# Statistical Trading Using Target Oriented Trading Agent

Zeeshan Ahmed

Vienna University of Technology,
Getreidemarkt 9/307 1060, Vienna, Austria
Mobile: 004369981302854, Email: zeeshan.ahmed@tuwien.ac.at
URL: www.tuwien.ac.at & www.zeeshanahmed.bravehost.com

*Abstract*— **In this article we briefly present our contributions toward Trading Agent Competition (TAC); an international forum for promotion of research into the trading agent problems. Moreover, we present some strategies proposed and used in the development of our TAC Agent and resultant brief information after its participation in a real time trading environment. In the end we conclude with needed improvements and future recommendations.**

## I. INTRODUCTION

TAC Trading Agent Competition is an international forum for the promotion of the research into the trading agent problems. TAC provides an excellent multi agent based research environment for autonomous software agent competition. It stimulates researchers and developers to design agents that can replace humans in rich competitive trading environment. It also provides an opportunity to find methods for best time interval calculation, best ask-price estimations, price predictions, needed calculations and utility maximizations. TAC Classic is a challenging market game in which agent competes with other agents aiming to buy travel packages for his client(s).

The uncertainty in hotel prices, preferences for scheduled flights and redundancy in entertainment tickets make this competition more interesting and challenging. There are 8 clients, 8 agents and a total of 28 auctions [1] as shown in Fig 1 .i.e.,

- 4 auctions for incoming flights
- 4 auctions for outgoing flights
- 4 auctions for better hotel rooms
- 4 auctions for alternative hotel rooms
- 4 auctions for 3 types of entertainments.

A complete package consists of a round trip including flight, hotel (Tampa and Shoreline Shanties) reservation, and ticket to some/any of available three entertainments. All clients have their own preferences for the trip e.g. arrival and departure days, type of hotels they wants to stay in, and entertainments. The target of each trading agent is to satisfy its customers by maximizing its utilities based on cost, penalties and bonuses [2]. Bonuses are of two kinds .i.e., Hotel, Fun. *Hotel Bonus* is the points awarded for each night in Tampa Towers which is a better hotel than Shoreline Shanties. *Fun Bonus* is sum of premiers received on entertainments.

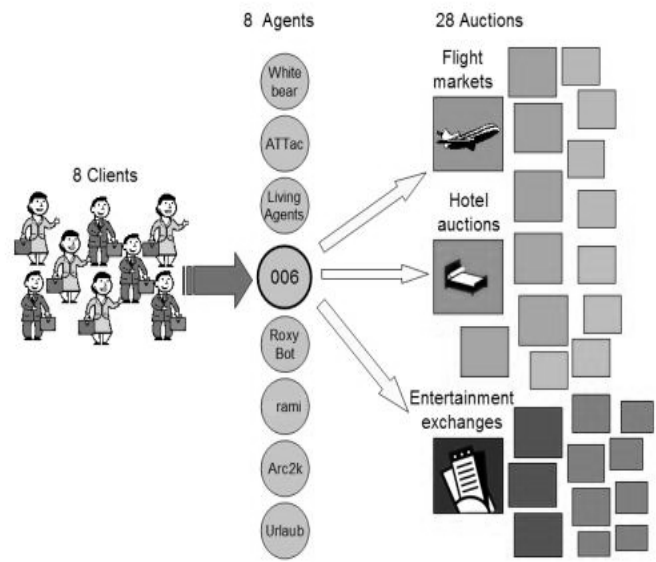

**Fig 1**. Trading Environment [1]

Whereas Travel Penalty is the points deducted for neglecting clients' preferences for arrival and departure dates. By close observation of conditions and analysis of formulae, one can easily find that Travel Penalty range from 0 to 600, Hotel Bonus from 0 to 150, and Fun Bonus from 0 to 600 which determines that the Utility can range from 400 to 1750 for each customer [1]. An agent can maximize its utility by maximizing Hotel Bonus and Fun Bonus, and minimizing Travel Penalty.

In this research paper we present our proposed and developed trading agent in section II. Moreover we present the result of its contribution in a real time multi agent trading environment in section III and describe future recommendations and improvements in section IV.

## II. TOTA

Keeping eyes on aforementioned requirements, problems and rules in mind we proposed and designed our own trading agent i.e., Target Oriented Trading Agent (TOTA). TOTA is an autonomous trading software agent programmed in Java [3]. TOTA is a designed and implemented to compete in the real time trading market considering the preferences of clients. Due to the dynamic nature of TAC Classic, there is a continuous change in prices and availability of items. It requires an agent to implement dynamic behavior inside and clever strategies for the maximization of utility.

Following strategies have been introduced and implemented in TOTA .i.e., Dynamic Planning, Travel Penalty Calculation, Utility Calculation, Best Time Intervals, Feasibility of Travel Packages, Auction Oriented Planning, Time-Dependent and Need-Based Bidding, Responding to the Competition and Selling the Redundant Tickets for Entertainment.

### A. Dynamic Planning

Due to the dynamic behavior of the competition a static agent who decides its plan at the start of game cannot survive in the competition. During the game, the prices of items may rise or fall or some time even becomes unavailable. TOTA therefore keeps an eye on dynamics of the environment and changes its plan according to the current situation of the market.

### B. Travel Penalty Calculation

TOTA calculates utility by taking the mod of difference of Arrival Date and Proffered Arrival Date values, mod of difference Departure Date and Proffered Departure Date values and then by multiplying with 100.

Equation to calculate Travel Penalty by TOTA is:

$$Travel\ Penalty = 100 * (|Arrival\ Date - Proffered\ Arrival\ Date|) + (|Departure\ Date - Proffered\ Departure\ Date|)$$

### C. Utility Calculation

TOTA calculates utility by adding Travel Penalty, Hotel Bonus, Fun Bonus and subtracting the whole from 1000. Equation to calculate utility by TOTA is:

$$Utility = 1000 - Travel\ Penalty + Hotel\ Bonus + Fun\ Bonus$$

### D. Best Time Intervals

Timing is an important factor to be considered in the dynamic trading environments. TOTA puts bids and updates its allocations on minute basis. We have calculated and observed 1 minute time interval as beast time interval because of random change in hotel auctions in one minute.

### E. Feasibility of Travel Packages

A feasible travel package is the one that includes all the required items and has a positive utility associated with it. TOTA switches hotel types if a package becomes infeasible. In the same way, it adjusts in-flights and out-flights to fit in the package

### F. Auction Oriented Planning

TOTA updates allocations every minute. When server closes its auctions, TOTA updates its plan according to the auctions.

### G. Time-Dependent and Need-Based Bidding

Flight prices start from 0 and keep on rising till the server closes auctions till the end of the game. It is therefore important for agents to put bids in time according to the needs. TOTA determines its needs for flights, updates its flight bids every 30 seconds and then places a bid after 8 minutes of the start of game.

### H. Responding to the Competition

Hotel prices increase with the passage of time. An agent can acquire a hotel by offering a bit higher price than others. For this reason, we need some intelligently calculated price to bid.

$$Price = (First\ Previous\ Bid - Second\ Previous\ Bid) + Ask\ Price$$

TOTA uses a very simple method to do this. It calculates the difference of the two previous bids and adds this in ask price to put bids.

### I. Selling the Redundant Tickets for Entertainment

According to the rule of trading, agent needs to sell the redundant entertainment tickets at the end of game because visiting the same event again in a tour will add nothing to the utility of client. While selling such tickets, we may be required to lower their price if we find no customer at some specific price. A random decrease in price is not a good strategy. TOTA lowers its selling price from 200 to 0 logarithmically with the passage of time.

## III. TOTA IN REAL TIME TRADING ENVIRONMENT

TOTA participated in the real time multi agent based trading competition organized by Blekinge Institute of Technology (BTH), The University of Blekinge Sweden [4]. Five games were played using eight agents each time. TOTA showed excellent performance in all the games.

## IV. FUTURE RECOMMENDATIONS & IMPROVEMENTS

There are some areas where TOTA needs to be improved in four areas to perform with more stability and reliability in real time trading environment, which are

1. Best time interval; finding best time intervals for putting bids and updating allocations.
2. Flight Shift In & Out; for flights, a better allocation algorithm can be introduced to shift in-flights and out-flights to best fit our needs.
3. Bid Price; better process to find bid price for hotels than using previous two values to find bid price for hotels.
4. Selling Ticket Price; Calculation of selling price for redundant entertainment tickets.

We noticed these areas of improvements, when we participated and tested TOTA in real time trading environment. We can improve these by introducing optimizations in used algorithms based on machine learning concepts e.g. learning and stochastic. A better approach can be devised to find the best ask price and enhance the maximization of the utility function.

## V. CONCLUSION

In this short article we have briefly presented Trading Agent Competition (TAC); an international forum for promotion of research into the trading agent problems. We have presented our contributions towards TAC by implementing a real time TAC Agent .i.e., TOTA. Moreover we have described our planed and used strategies in the development of TAC Agent. In the end we have concluded with the needed improvements and future recommendations for TOTA.

## VI. REFERENCES

[1] Trading Agent Competition.– TAC Classic – Game Description, http://www.sics.se/tac/page.php?id=3
[2] Joakim Eriksson and Sverker Janson, "The Trading Agent Competition – TAC 2002", ERCIM News No. 51, October 2002
[3] Java, Sun Microsystems, reviewed on 20 March 2008 < www.java.sun.com/>
[4] Blekinge Institute of Technology (BTH), reviewed on 20 March 208 <www.bth.se >